## Research Article
# Forecasting Drought Using Multilayer Perceptron Artificial Neural Network Model

**Zulifqar Ali,[1] Ijaz Hussain,[1] Muhammad Faisal,[2,3] Hafiza Mamona Nazir,[1] Tajammal Hussain,[4] Muhammad Yousaf Shad,[1] Alaa Mohamd Shoukry,[5,6] and Showkat Hussain Gani[7]**

[1]*Department of Statistics, Quaid-i-Azam University, Islamabad, Pakistan*
[2]*Faculty of Health Studies, University of Bradford, Bradford BD7 1DP, UK*
[3]*Bradford Institute for Health Research, Bradford Teaching Hospitals NHS Foundation Trust, Bradford, UK*
[4]*Department of Statistics, COMSATS Institute of Information Technology, Lahore, Pakistan*
[5]*Arriyadh Community College, King Saud University, Riyadh, Saudi Arabia*
[6]*Workers University, Cairo, Egypt*
[7]*College of Business Administration, King Saud University, Muzahimiyah, Saudi Arabia*

Correspondence should be addressed to Ijaz Hussain; ijaz@qau.edu.pk

These days human beings are facing many environmental challenges due to frequently occurring drought hazards. It may have an effect on the country's environment, the community, and industries. Several adverse impacts of drought hazard are continued in Pakistan, including other hazards. However, early measurement and detection of drought can provide guidance to water resources management for employing drought mitigation policies. In this paper, we used a multilayer perceptron neural network (MLPNN) algorithm for drought forecasting. We applied and tested MLPNN algorithm on monthly time series data of *Standardized Precipitation Evapotranspiration Index (SPEI)* for seventeen climatological stations located in Northern Area and KPK (Pakistan). We found that MLPNN has potential capability for SPEI drought forecasting based on performance measures (i.e., Mean Average Error (MAE), the coefficient of correlation ($R$), and Root Mean Square Error (RMSE)). Water resources and management planner can take necessary action in advance (e.g., in water scarcity areas) by using MLPNN model as part of their decision-making.

## 1. Introduction

The demand of water has increased diversely due to expansion in agriculture, population, energy, and industrial zone. Many parts of the world suffered each year due to scarcity of water. Change in climatic condition and contamination in water play a key role in water scarcity, Aswathanarayana [1].

Drought can be recognized as disaster associated with climate that can have effect on a wide range of land. There are many factors that play a major role in drought occurrence including high wind, low relative humidity, temperature, and characteristics and duration of rain, intensity, and onset, Wilhite [2]. Drought can be one of the main sources in reducing freshwater flows and has huge impact on the planning and management of water resources.

Several tools have been used for the assessment of drought. Drought indices are one of the most commonly used tools for assessing the drought conditions around the world and few of them are as follows: Rainfall Anomaly Index (RAI), Van Rooy [3] and Decile Gibbs [4]; Crop Moisture Index (CMI), Palmer [5]; the Palmer Drought Severity Index (PDSI), Palmer [6]; Bhalme and Mooly Index (BMI), Bhalme and Mooley [7]; Surface Water Supply Index (SWSI), Shafer and Dezman [8]; Reclamation Drought Index (RDI), Weghorst [9]; Standardized Precipitation Index (SPI), McKee et al. [10]; and Standardized Precipitation Evapotranspiration Index (SPEI), Vicente-Serrano et al. [11]. Drought indices are efficient tools instead of making decision on raw data. In this study, we reviewed these drought indices to understand the appropriateness of each drought index.

Similar to drought assessment tools, several models have been developed for drought forecasting. Paulo and Pereira [12] applied Markov chain on *SPI* to characterize the stochasticity of drought and predict three months ahead drought class. Neural network is an information processing method, which adaptively determine pattern from data. Hypothetically, it has been exposed that, given a suitable number of nonlinear processing units, neural network can learn from, practice, and calculate approximately any complex function with greater accuracy [13, 14]. Kim and Valdés [15] forecasted drought using dyadic wavelet transforms and neural network. Mishra et al. [16] used *SPI* to compare the forecasting performance of Artificial Neural Network (ANN) and linear stochastic model in the Kangsabati River Basin, India. Bacanli et al. [17] investigated SPI and used Adaptive Neurofuzzy Inference System (ANFIS) for drought forecasting.

A few applications of ANN models in drought forecasting only comprised of Morid et al. [18]. Mishra and Desai [19] compared linear stochastic models (e.g., Autoregressive Integrated Moving Average (ARIMA), seasonal ARIMA, Recursive Multistep Neural Network (RMSNN), and Direct Multistep Neural Network (DMSNN)) for drought forecasting using SPI time series data of Kangsabati River Basin. They found that DMSNN is helpful in long-term drought forecasting; however, RMSNN is useful in short-term drought forecasting.

The log linear model is class of generalized linear models that can explore the relationships among categorical variables, Agresti [20]. Moreira et al. [21] used three-dimensional log linear model for drought forecasting and found it is a useful tool for temporary drought warning systems. Morid et al. [18] used ANN to predict the values of two drought indices, SPI and Effective Drought Index (EDI). A variety of different structures of ANNs were applied on SPI and EDI time series data with different time scale for several stations of Tehran (Iran). Both indices have $R^2$ values within 0.66–0.79 for 6-month time scale. However, it was shown that the EDI forecasted results were better to those of SPI in all lead times. Marj and Meijerink [22] purposed a model for forecasting agricultural drought based on Normalized Difference Vegetation Index (NDVI) by using Effective Climate Signal (EDS) and ANN approach. These models were applied at Ahar-Chay Basin in Azerbaijan Province. Their results show that, in spring, synthetic NDVI can be forecasted using ANN's approach. Fathabadi et al. [23] used time series data of SPI and applied ANN and *K*-Nearest Neighbor (KNN) models to forecast drought at five stations of Western Iran. Their results show that ANNs perform better than KNN for forecasting SPI values for 9- and 12-month time scale.

Conventionally, hydrological variables, like monthly precipitation and temperature, have been widely modeled using different linear techniques, such as Autoregressive Moving Average (ARMA) Salas and Boes [24] and Seasonal Autoregressive Integrated Moving Average (SARIMA), Mishra and Desai [19]. The ANNs have showed outstanding ability in modeling and forecasting nonlinear and nonstationary time series data in water resources and hydrology, Goovaerts [25]. This main feature of ANN makes it an attractive method for drought forecasting, Morid et al. [18]. In recent years, due to this advantage, many researchers have applied ANN modeling approach in different fields [16, 18, 19, 21, 26].

In this study, due to the importance of drought forecasting, the capability of *ANN* model is evaluated by forecasting drought using *multiscalar drought Index-SPEI* at various climatic zones of Pakistan. The rest of the paper is organized as follows. The brief description about spatial domain and estimation method for SPEI are presented in Section 2. The neural network model for forecasting the drought index and its testing and validation are presented in Section 3. Finally, we concluded our results in Section 4.

## 2. Material and Methods

*2.1. Study Area.* Our study area is in Northern Area and KPK including capital territory of Pakistan. We collected monthly data on total rainfall and mean temperature from seventeen meteorological stations (Balakot, Kotli, Cherat, Chilas, Islamabad, Gupis, Peshawar, Saidu Sharif, Muzaffarabad, Bunji, DI Khan, Drosh, Garhi Dupatta, Dir, Gilgit, and Kakul) from 1975 to 2012. As these stations' data are managed by the Pakistan Meteorological Department (PMD), Islamabad, we collected the data from the Karachi Data Processing Center via PMD. The selected locations represent fully precipitation regimes affecting the area where water is the main source for agriculture and hydropower for the flood plains in Pakistan. These stations have significant ecological role, including watershed and enhancing the lifespan of Tarbela Dam. This dataset contains catchments with minimum synthetic influences and have good hydrometric performance. In this paper, SPEI with four different time scales are calculated for each station.

*2.2. Standardized Precipitation Evapotranspiration Index (SPEI).* Vicente-Serrano et al. [11] developed a new multiscalar drought index called SPEI, which is based on both temperature and precipitation data. The SPEI is an extension of the extensively used drought index called SPI. The SPEI is proposed to report both precipitation and Potential Evapotranspiration (PET) in determining drought.

Different equations are used to estimate PET values according to the nature of data that linked PET values with temperature data. The most commonly used procedures for calculating PET are Thornthwaite equation, Thornthwaite [29]; Penman equation, Allen et al. [30] and Allen and Pruitt [31]. In this study, Thornthwaite equation is used to estimate PET values. Estimation procedures for SPEI and SPI are similar. SPI uses only time series data of precipitation, recorded with different time scale as an input. However, the SPEI uses time series data on both precipitation and temperature. The procedure for estimation of SPEI is as follows:

$$\text{PET} = 16k\left(\frac{10T}{i}\right)^m. \quad (1)$$

In the above equation, $T$ is monthly temperature in degree Celsius and $i$ is heat index derived from 12-month index

Table 1: The SPEI drought category classification provided by McKee et al. [10].

| SPEI values | Drought classes |
|---|---|
| ≥2 | Extremely wet |
| 1.5 to 1.99 | Very wet |
| 1 to 1.49 | Moderate wet |
| .99 to −.99 | Near normal |
| −1 to −1.49 | Moderate drought |
| −1.5 to 1.99 | Severe drought |
| ≤−2 | Extreme drought |

values calculated as a sum of 12-month index values $i$, which is calculated as follows:

$$i = \left(\frac{T}{5}\right)^{1.514}. \tag{2}$$

$m$ is a coefficient depending on $i$, and $k$ is a correction coefficient computed as a function of the latitude and month. The difference between precipitation and PET provides a measure of water surplus or deficit for the month and this is compared over time and standardized to get the value of SPEI.

$$d_i = P_i - \text{PET}. \tag{3}$$

SPEI values were obtained by fitting the long-term record of difference between precipitation and PET for specified time interval of any location.

Vicente-Serrano et al. [11] used same classification criteria of drought as described by McKee et al. [10]. Table 1 shows the classification of SPEI values corresponding with climatic classes provided by McKee et al.

Sönmez et al. [32] used the Gamma distribution to investigate spatiotemporal variability in meteorological droughts at Turkey.

Mathematically, the SPEI is based on the cumulative probability distribution function of a given quantitative values of rainfall occurrence for a specific station.

In this study, we calculate SPEI values by standardizing different probability distributions (e.g., Gamma, Generalized Extreme Values Distribution, Log-Logistic Distribution, and Generalized Pareto Distribution) that fit the $d_i$ time series. Kolmogorov-Smirnov test, Justel et al. [33], and Anderson Darling test, Anderson and Darling [34], for goodness of measure are applied using Easy-Fit, Schittkowski [35], computer application before standardizing the most appropriate distribution. Detailed discussion on these goodness-of-fit tests is skipped in this section.

McKee et al. [10] transformed Gamma distribution into a normal distribution by using inverse normal (Gaussian) function in order to calculate SPI values. To estimate the parameters of each distribution that fit well, different methods of parameter estimation are used. Table 2 shows probability distributions corresponding to the estimation method of parameters for each distribution.

The resulting parameters of each distribution are then used to derive Cumulative Distribution Function (CDF). For undefined values of $x$, for example, in case of the Gamma distribution, the rainfall time series data may contain zero rainfall. The cumulative distribution of zero and nonzero rainfall is calculated by the following expression:

$$H(x) = q + (1 - q) F(x), \tag{4}$$

where $q$ is the probability of zero rainfall.

Table 2: Parameter estimation method for different distribution.

| Probability distribution | Method of estimation |
|---|---|
| Gamma | Method of moments |
| Generalized Pareto | Method of L-moments |
| Generalized Pareto | Method of L-moments |
| Generalized Extreme Value | Method of L-moments |
| Generalized Extreme Value | Maximum likelihood method |
| Log-Logistic | Method of moments |

If $m$ is the number of zeros present in a rainfall time series data, then $q$ is estimated by $m/n$.

The distribution function of each probability distribution is than transformed into standard normal distribution to obtain SPTI values having zero mean and unit variance.

Following Mishra and Desai and McKee et al., the current study employed the approximate transformation provided by Abramowitz and Stegun [36] to transform the cumulative probability distribution into a standardized normal distribution, which are given as follows:

$$\text{SPEI} = -\left(k + \frac{c_o + c_1 k + c_c k^2}{1 + d_1 k + d_2 k^2 + d_3 k^3}\right) \tag{5}$$

for

$$k = \sqrt{\ln\left(\frac{1}{(H(x))^2}\right)} \tag{6}$$

when

$$0 < H(X) \le 0.5,$$

$$\text{SPEI} = +\left(k - \frac{c_o + c_1 k + c_c k^2}{1 + d_1 k + d_2 k^2 + d_3 k^3}\right) \tag{7}$$

and for

$$k = \sqrt{\ln\left(\frac{1}{(1 - H(x))^2}\right)} \tag{8}$$

when

$$0.5 < H(X) \le 1, \tag{9}$$

where

$$c_o = 2.515517,$$
$$c_1 = 0.802853,$$
$$c_2 = 0.010328,$$
$$d_1 = 1.432788,$$

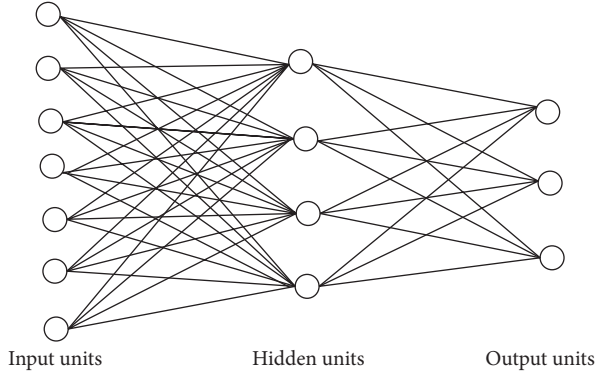

Figure 1: An example of a simple feedforward network, Garson [27].

$$d_2 = 0.985269,$$
$$d_3 = 0.001308. \tag{10}$$

The average value of the SPEI is 0, and the standard deviation is 1. The SPEI is a standardized variable; therefore, it can be compared with other SPEI values over time and space. The SPEI value equal to 0 indicates a value corresponding to 50% of the cumulative probability of $D$, according to a Log-Logistic Distribution.

## 3. Neural Network Forecasting

There are several methods for the development and implementation of neural network model of forecasting. In many applications, feedforward neural network topology with backpropagation learning algorithm was used, while some used variant of this. Several researchers described the problem in finding the appropriate network size for predicting real-world time series, Zhang et al. [37].

The MLPNN model is the most extensively used type of ANN's approach for modeling hydrological data, Wang et al. [38]. MLP model belongs to a general class structure of ANN called feedforward neural network. A feedforward neural network is a basic type of neural network that is capable of approximating both continuous and integrable functions. Network architecture of MLP consists of neurons that grouped in layers.

In MLPNN model, all the input nodes are in one layer and hidden layer is distributed into one or more hidden layers. Figure 1 shows a general structure of simple feedforward network.

Suppose there are $N$ layers in MLP: first layer is called input, $N$th layer is the output, and 2 to $N-1$ layers are hidden layers. Assume that there are $L_l$ neurons, where, $L_l = 1, 2, 3, \ldots, N$.

Let $w_{ij}^L$ and $x_{ij}$ be the weight and $i$th be the neuron, respectively, such that $1 \leq j \leq L_{n-1}$, $1 \leq i \leq L_n$, where $w_{ij}$ are the weights and $x_{ij}$ is the external input for model.

Let $Z_i$ be the output of the $i$th neuron of $N$th layer.

Also, let $w_{io}^n$ be the extra weight parameter that represent bias of $i$th neuron of $N$th layer such that $w$ includes $w_{ij}^n$.

That is

$$w = \left[w_{ij}^1, w_{ij}^2, w_{ij}^3, \ldots, w_{L_N L_{N-1}}^N\right], \tag{11}$$

where

$$\begin{aligned} j &= 0, 1, 2, \ldots, L_{n-1}, \\ i &= 1, 2, 3, \ldots, L, \\ n &= 1, 2, 3, \ldots, N. \end{aligned} \tag{12}$$

For designing ANN architecture, one must determine the optimum number of the following layers:

(i) The number of input layers
(ii) The number of hidden layers
(iii) The number of output layers

Figure 2 illustrates general architecture of MLPNN model.

In this research, MLPNN model of ANNs was used for drought forecasting.

Detailed explanation MLPNN model and its selection of parameters is given in the following section.

*3.1. Multilayer Perception Architecture.* All neural networks have an input layer and an output layer; however, the number of hidden layers may vary. Basically, selection of these variables is domain-specific or depends on the problem. Many algorithms, such as the polynomial time algorithm, Roy et al. [39]; the pruning algorithm, Sietsma and Dow [40]; the canonical decomposition technique, Wang et al. [41]; and network information criterion, Murata et al. [26], have been proposed to find optimum structure of the network, but none of these methods guaranteed the optimal solution of the parameters for all types of forecasting problems.

Literature shows that there is no systematic way to investigate these problems. Many researchers adopted trial and error methodology for a specific problem which is the basic cause of inconsistency in ANN literature, Sheela and Deepa [42]. Zhang et al. [37] reported that there is not any structured model that identify which network structure would be the best. There are no hard and fast rules prevailing the correct structure of a neural network. Important factors such as the number of inputs, the number of hidden units, and the arrangement of these units into layers are often determined using trial and error methods or fixed in advance according to the subjective opinion of each individual designer, Fischer and Gopal [43].

The procedure for MLPNN consists of four parts:

(i) Variable selection
(ii) Formulations of training, testing, and validation
(iii) Architecture
(iv) Model verification and forecasting

Current research employed the MLPNN model by using Zaitun time series software. Following Lipae et al. [44], the selection of variable is based on software. This employs that choosing variable for training, testing, and validation is done by software itself. Application of same methodological structure can be found in Babić et al. [45]; Lipae et al. [44]; Kadiyala and Kumar [46]. However, the risk of overfitting of the MLPNN was taken by early stopping condition. As we used iterative method for training a learner, this stopping condition fits better the data with each iteration. There are two basic rules of stopping condition (i.e., mean square error value and mean square error change). These rules help give guidance about the number of iterations running before the initialization of overfitting of learner, Prechelt [47]. In Zaitun time series software, one can find stopping condition in neural network analysis form. Several models are applied and tested with various combinations of layers (i.e., input layer, hidden layer, and output layer) and four activation functions (i.e., semilinear, sigmoid bipolar sigmoid, and the hyperbolic tangent function). Following Gowda and Mayya [48], the parameters of the ANN architecture in terms of learning rate, momentum, bias, the number of hidden neurons, and the activation constant were considered. Trial and error procedure was adopted to choose the optimal value of each structured parameter of network model. The developed ANN model consists of 3 layers that are input, hidden, and output of 30 neurons, 8 neurons, and 1 neuron, respectively. For verification of forecast model, the residuals series were tested and plotted to examine whether the series is uncorrelated or not. If the residuals revealed to be uncorrelated, the selected model is then applied to forecast drought indices. We found that sigmoid function is best for each drought index for one-month scale data based on the criterion of mean square error. Momentum of 0.5 and training epoch of 10000 were set. The input vector consists of previous 30 values of each index. After selection of the appropriate parameter of ANN, the forecasted model for each of the indices is then validated on 20% of data. Validation of forecasted model is done based on performance measures: MAE, the correlation coefficient ($R$), and RMSE.

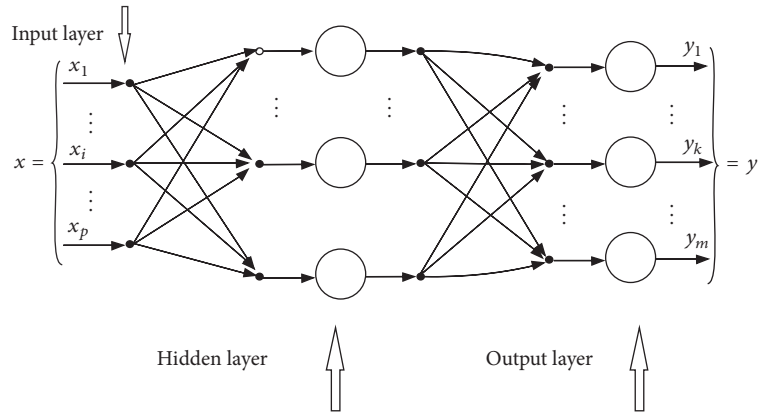

Figure 2: General architecture of multilayer perceptron neural network model, Sherrod [28].

### 3.2. Results and Discussion.
In this study, time series data on observed SPEI with different time scale are computed by standardizing the probability distribution that describes well behavior of difference between precipitation and evapotranspiration using Abramowitz and Stegun [36] approach. After computing the drought indices, the multilayer perceptron model was used to describe the method of forecasting the quantitative values of SPEI for each selected stations of our study area. Each forecasted model was evaluated by considering the correlation coefficient ($R$), RMSE, and MAE. Time series data of observed SPEI with 1-, 3-, 6-, and 12-month time scale at selected stations are compared with predicted values of SPEI.

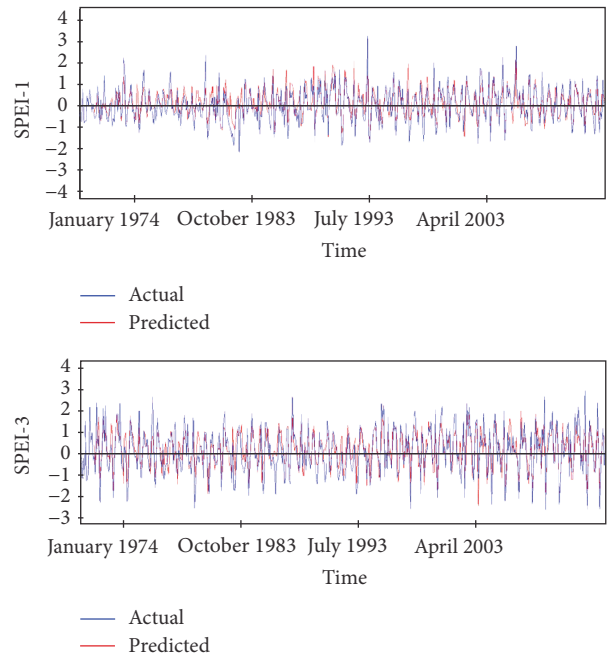

Figure 3: The observed and predicted values of SPEI using multilayer perceptron in the validation phase for Balakot station.

Figures 3, 4, 5, and 6 show graphical representation of historical observed values and predicted values of SPEI with different time scale of stations (e.g., Astor and Balakot). Table 3 shows results of MLPNN model summaries for 1-, 3-, 6-, and 12-month time scale SPEI values of each study station.

The model is potentially able to predict drought condition by using SPEI values with different time scale. The excellence of the forecast is reflected in the correlation coefficient between observed and estimated time series, the RMSE and MAE.

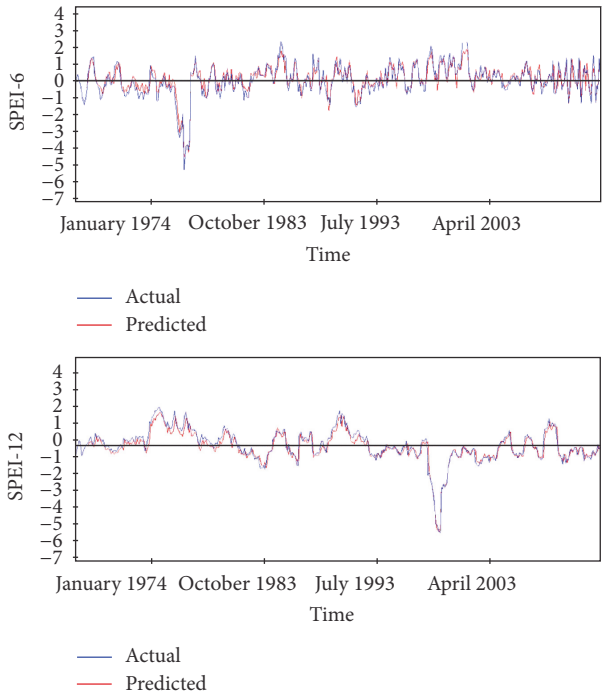

Figure 4: The observed and predicted values of SPEI using multilayer perceptron in the validation phase for Balakot station.

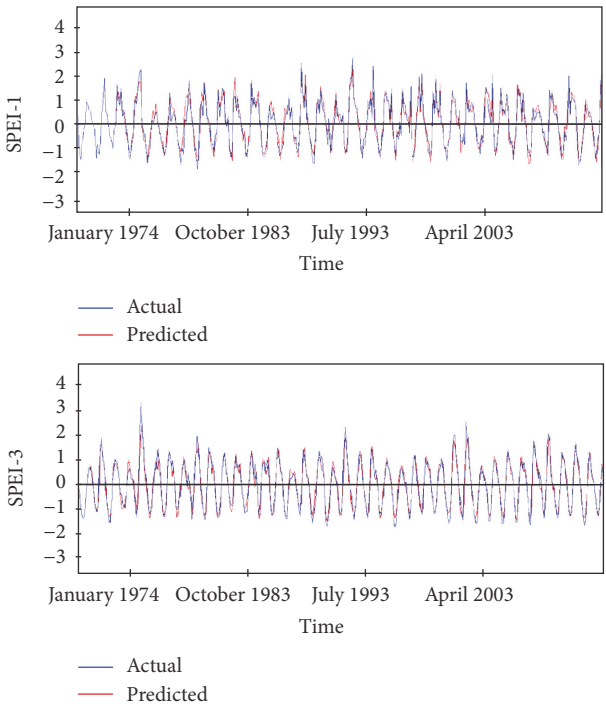

Figure 5: The observed and predicted values of SPEI using multilayer perceptron in the validation phase for Astor station.

The accuracy of the selected model in all stations for each index is good in terms of correlation between observed and predicted SPEI values. From each observatory, correlation coefficient ranges lie in the interval 0.887 to 0.987 for SPEI-1, 0.876 to 0.994 for SPEI-3, 0.876 to 0.994 for SPEI-6, and 0.780 to 0.970 for SPEI-12. Figure 7 shows actual and forecasted plots for stations (e.g., Astor, Balakot, Chilas, and Dir). Results show that ANN modeling works extremely well for forecasting SPEI values with different time scale; however, for Balakot station, deviations are apparent between observed and predicted values of SPEI-12 and correlation coefficient is 0.780.

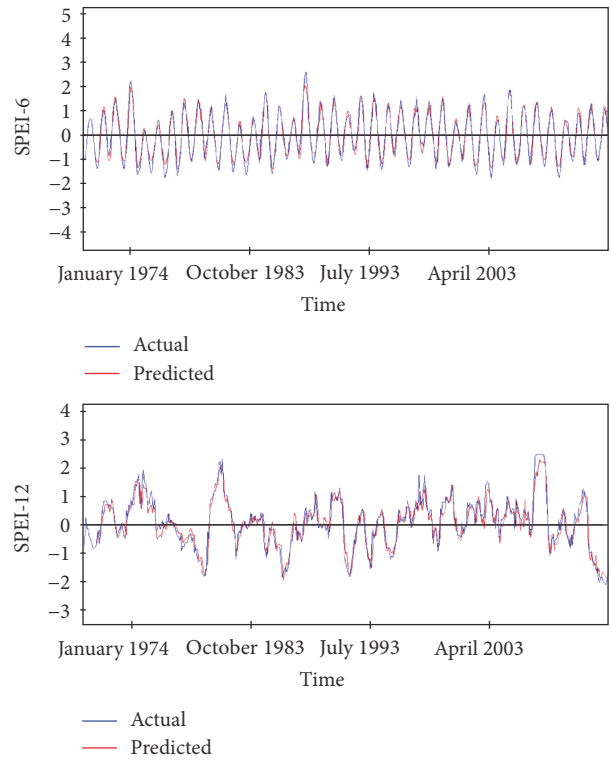

Figure 6: The observed and predicted values of SPEI using multilayer perceptron in the validation phase for Astor station.

By using forecasted quantities of drought indices as input, model can be used to predict further by a multistep approach. No evidence was found about significant deviation between observed and predicted values of drought index for all indices.

## 4. Conclusion

In this study, multilayer perceptron neural network (MLPNN) algorithm is used for nonlinear drought forecasting of monthly time series data of average temperature and total precipitation that recorded from seventeen synoptic stations of Northern Area and KPK (Pakistan) from 1975 to 2012. SPEI values were estimated by fitting appropriate probability distribution of difference between precipitation and PET. We found that the MLPNN model is convenient for operational purposes (i.e., water resources and management) as variation between input data of observed and predicted SPEI values is not high.

TABLE 3: The performance for all study stations of multilayer perceptron models in validation phase.

| Stations | SPEI-1 | | | SPEI-3 | | | SPEI-6 | | | SPEI-12 | | |
|---|---|---|---|---|---|---|---|---|---|---|---|---|
| | MSE | MAE | R | MSE | MAE | R | MSE | MAE | R | MSE | MAE | R |
| Kakul | 0.107 | 0.254 | 0.946 | 0.254 | 00.22 | 0.916 | 0.256 | 0.254 | 0.886 | 0.263 | 0.632 | 0.886 |
| Astor | 0.100 | 0.241 | 0.949 | 0.089 | 0.248 | 0.947 | 0.094 | 0.741 | 0.947 | 0.258 | 0.263 | 0.925 |
| DI Khan | 0.078 | 0.205 | 0.947 | 0.099 | 0.220 | 0.927 | 0.094 | 0.263 | 0.927 | 0.096 | 0.523 | 0.963 |
| Balakot | 0.181 | 0.339 | 0.921 | 0.098 | 0.246 | 0.934 | 0.093 | 0.749 | 0.944 | 0.0589 | 0.236 | 0.780 |
| Bunji | 0.099 | 0.217 | 0.944 | 0.099 | 0.230 | 0.914 | 0.091 | 0.746 | 0.944 | 0.089 | 0.856 | 0.942 |
| Chilas | 0.100 | 0.244 | 0.946 | 0.104 | 0.207 | 0.941 | 0.121 | 0.785 | 0.941 | 0.123 | 0.856 | 0.936 |
| Dir | 0.099 | 0.250 | 0.946 | 0.100 | 0.245 | 0.954 | 0.421 | 0.259 | 0.994 | 0.125 | 0.456 | 0.926 |
| Drosh | 0.100 | 0.242 | 0.942 | 0.099 | 0.230 | 0.944 | 0.062 | 0.230 | 0.944 | 0.096 | 0.236 | 0.985 |
| Garhi Dupatta | 0.099 | 0.251 | 0.945 | 0.099 | 0.243 | 0.876 | 0.290 | 0.245 | 0.876 | 0.091 | 0.226 | 0.872 |
| Kotli | 0.121 | 0.258 | 0.940 | 0.100 | 0.239 | 0.948 | 0.123 | 0.485 | 0.948 | 0.105 | 0.256 | 0.923 |
| Cherat | 0.175 | 0.321 | 0.917 | 0.100 | 0.224 | 0.939 | 0.101 | 0.785 | 0.939 | 0.103 | 0.845 | 0.926 |
| Islamabad | 0.182 | 0.327 | 0.911 | 0.098 | 0.231 | 0.921 | 0.091 | 0.846 | 0.921 | 0.094 | 0.785 | 0.952 |
| Peshawar | 0.022 | 0.109 | 0.987 | 0.099 | 0.222 | 0.934 | 0.093 | 0.856 | 0.934 | 0.097 | 0.159 | 0.942 |
| Muzaffarabad | 0.102 | 0.238 | 0.950 | 0.099 | 0.248 | 0.994 | 0.098 | 0.286 | 0.984 | 0.09\3 | 0.741 | 0.964 |
| Gilgit | 0.187 | 0.332 | 0.887 | 0.140 | 0.275 | 0.900 | 0.145 | 0.869 | 0.900 | 0.126 | 0.451 | 0.970 |
| Gupis | 0.027 | 0.126 | 0.985 | 0.097 | 0.225 | 0.930 | 0.045 | 0.856 | 0.930 | 0.087 | 0.236 | 0.910 |
| Saidu Sharif | 0.100 | 0.244 | 0.945 | 0.100 | 0.243 | 0.944 | 0.145 | 0.265 | 0.944 | 0.115 | 0.263 | 0.894 |

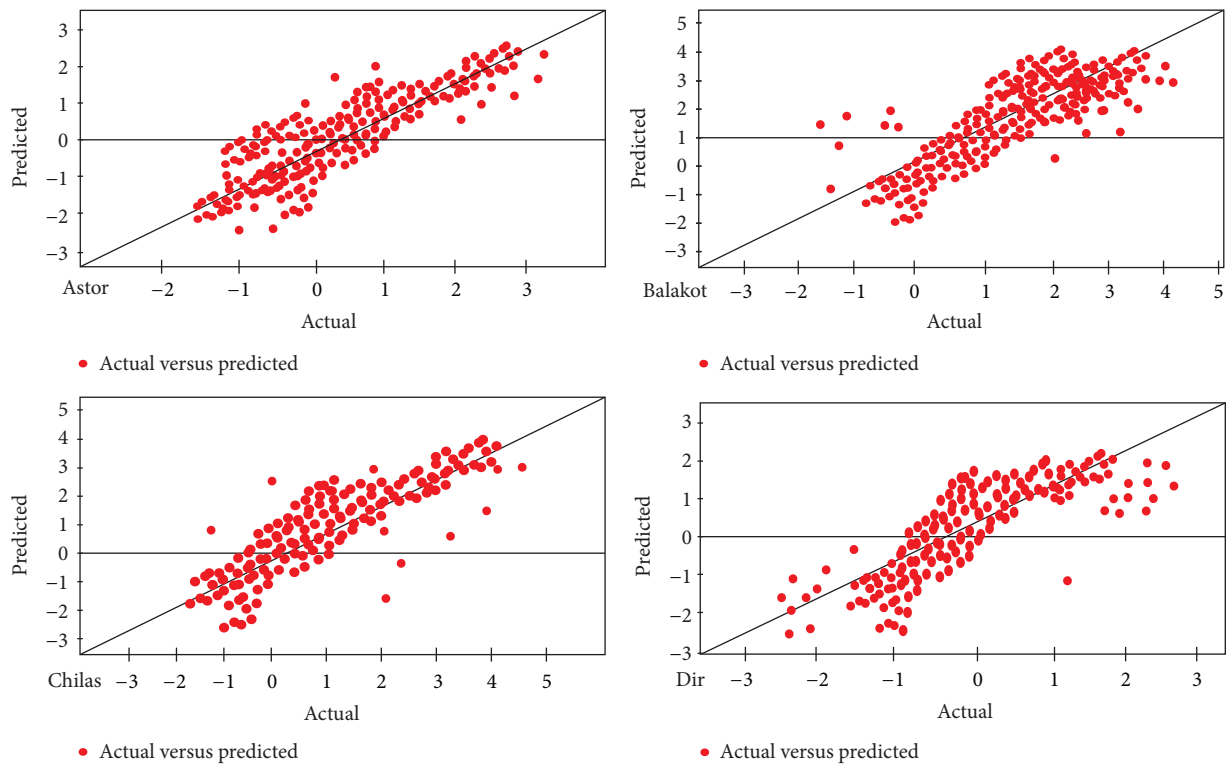

FIGURE 7: Actual versus predicted graph of SPEI-1.

Outcomes associated with the study show that ANNs have the power to capture the variation in selected drought indices with one-month time scale. Water resources and management planner may take help from the developed neural network model to take action in advance to know about where water scarcity is increasing owing to insufficient rainfall in a particular region that may lead to drought condition.

## Ethical Approval

The manuscript is prepared in accordance with the ethical standards of the responsible committee on human experimentation and with the latest version (2008) of Helsinki Declaration of 1975.

## Disclosure

The manuscript is prepared by using secondary data.

## Conflicts of Interest

The authors declared that there are no conflicts of interest.

## Acknowledgments

The authors extend their appreciation to the Deanship of Scientific Research at King Saud University for funding this work through Research Group no. RG-1437-027.

## References

[1] U. Aswathanarayana, *Water Resources Management and the Environment*, CRC Press, 2001.

[2] D. A. Wilhite, *Preparing for Drought: A Guidebook for Developing Countries*, Diane Publishing, 1994.

[3] M. Van Rooy, "A rainfall anomaly index independent of time and space," *Notos*, vol. 14, pp. 43–48, 1965.

[4] W. J. Gibbs, "Rainfall deciles as drought indicators," 1967.

[5] W. C. Palmer, "Keeping track of crop moisture conditions, nationwide: The new crop moisture index," 1968.

[6] W. C. Palmer, *Meteorological Drought*, US Department of Commerce, Weather Bureau, Washington, DC, USA, 1965.

[7] H. N. Bhalme and D. A. Mooley, "Large-scale droughts/ floods and monsoon circulation.," *Monthly Weather Review*, vol. 108, no. 8, pp. 1197–1211, 1980.

[8] B. A. Shafer and L. E. Dezman, "Development of a surface water supply index (swsi) to assess the severity of drought conditions in snowpack runoff areas," in *Proceedings of the Western Snow Conference*, vol. 50, pp. 164–175, 1982.

[9] K. Weghorst, "The reclamation drought index: Guidelines and practical applications," in *proceedings of the North American Water and Environment Congress & Destructive Water*, pp. 637–642, ASCE, 1996.

[10] T. B. McKee, N. J. Doesken, and J. Kleist, "The relationship of drought frequency and duration to time scales," in *Proceedings of the 8th Conference on Applied Climatology*, vol. 17, pp. 179–183, American Meteorological Society, Boston, Mass, USA, 1993.

[11] S. M. Vicente-Serrano, S. Beguería, and J. I. López-Moreno, "A multiscalar drought index sensitive to global warming: the standardized precipitation evapotranspiration index," *Journal of Climate*, vol. 23, no. 7, pp. 1696–1718, 2010.

[12] A. A. Paulo and L. S. Pereira, "Prediction of SPI drought class transitions using Markov chains," *Water Resources Management*, vol. 21, no. 10, pp. 1813–1827, 2007.

[13] M. H. Hassoun, *Fundamentals of Artificial Neural Networks*, MIT press, 1995.

[14] R. S. Govindaraju and A. R. Rao, *Artificial Neural Networks in Hydrology*, Springer Publishing Company, Incorporated, 2010.

[15] T. W. Kim and J. B. Valdés, "Nonlinear model for drought forecasting based on a conjunction of wavelet transforms and neural networks," *Journal of Hydrologic Engineering*, vol. 8, no. 6, pp. 319–328, 2003.

[16] A. K. Mishra, V. R. Desai, and V. P. Singh, "Drought forecasting using a hybrid stochastic and neural network model," *Journal of Hydrologic Engineering*, vol. 12, no. 6, pp. 626–638, 2007.

[17] U. G. Bacanli, M. Firat, and F. Dikbas, "Adaptive Neuro-Fuzzy inference system for drought forecasting," *Stochastic Environmental Research and Risk Assessment*, vol. 23, no. 8, pp. 1143–1154, 2009.

[18] S. Morid, V. Smakhtin, and K. Bagherzadeh, "Drought forecasting using artificial neural networks and time series of drought indices," *International Journal of Climatology*, vol. 27, no. 15, pp. 2103–2111, 2007.

[19] A. K. Mishra and V. R. Desai, "Drought forecasting using feed-forward recursive neural network," *Ecological Modelling*, vol. 198, no. 1, pp. 127–138, 2006.

[20] A. Agresti, *An Introduction to Categorical Data Analysis*, vol. 423, John Wiley & Sons, 2007.

[21] E. E. Moreira, C. A. Coelho, A. A. Paulo, L. S. Pereira, and J. T. Mexia, "Spi-based drought category prediction using loglinear models," *Journal of Hydrology*, vol. 354, no. 1, pp. 116–130, 2008.

[22] A. F. Marj and A. M. J. Meijerink, "Agricultural drought forecasting using satellite images, climate indices and artificial neural network," *International Journal of Remote Sensing*, vol. 32, no. 24, pp. 9707–9719, 2011.

[23] A. Fathabadi, H. Gholami, A. Salajeghe, H. Azanivand, and H. Khosravi, "Drought forecasting using neural network and stochastic models," *Advances in Natural & Applied Sciences*, vol. 3, no. 2, 2009.

[24] J. D. Salas, *Applied modeling of hydrologic time series*, Water Resources Publication, 1980.

[25] P. Goovaerts, "Geostatistical approaches for incorporating elevation into the spatial interpolation of rainfall," *Journal of Hydrology*, vol. 228, no. 1-2, pp. 113–129, 2000.

[26] N. Murata, S. Yoshizawa, and S.-I. Amari, "Network Information Criterion-Determining the Number of Hidden Units for an Artificial Neural Network Model," *IEEE Transactions on Neural Networks*, vol. 5, no. 6, pp. 865–872, 1994.

[27] J. Garson, "Connectionism," in *The Stanford Encyclopedia of Philosophy*, E. N. Zalta, Ed., Spring, 2015.

[28] P. H. Sherrod, "Dtreg predictive modeling software," 2003 http://www.dtreg.com.

[29] C. W. Thornthwaite, "An approach toward a rational classification of climate," *Geographical Review*, vol. 38, no. 1, pp. 55–94, 1948.

[30] R. G. Allen, L. S. Pereira, D. Raes, M. Smith et al., "Crop evapotranspiration-guidelines for computing crop water requirements-fao irrigation and drainage paper 56," *FAO, Rome*, vol. 300, no. 9, article 6541, 1998.

[31] R. G. Allen and W. O. Pruitt, "Rational use of the FAO Blaney-Criddle formula," *Journal of Irrigation and Drainage Engineering*, vol. 112, no. 2, pp. 139–155, 1986.

[32] F. K. Sönmez, A. Ü. Kömüscü, A. Erkan, and E. Turgu, "An analysis of spatial and temporal dimension of drought vulnerability in Turkey using the standardized precipitation index," *Natural Hazards*, vol. 35, no. 2, pp. 243–264, 2005.

[33] A. Justel, D. Peña, and R. Zamar, "A multivariate kolmogorov-smirnov test of goodness of fit," *Statistics & Probability Letters*, vol. 35, no. 3, pp. 251–259, 1997.

[34] T. W. Anderson and D. A. Darling, "Asymptotic theory of certain "goodness of fit" criteria based on stochastic processes," *The Annals of Mathematical Statistics*, pp. 193–212, 1952.

[35] K. Schittkowski, "EASY-FIT: A software system for data fitting in dynamical systems," *Structural and Multidisciplinary Optimization*, vol. 23, no. 2, pp. 153–169, 2002.

[36] M. Abramowitz and I. A. Stegun, *Handbook of Mathematical Functions: with Formulas, Graphs, and Mathematical Tables*, Courier Dover Publications, 2012.

[37] G. Zhang, B. Eddy Patuwo, and M. Y. Hu, "Forecasting with artificial neural networks: the state of the art," *International Journal of Forecasting*, vol. 14, no. 1, pp. 35–62, 1998.

[38] W. Wang, P. H. Van Gelder, J. Vrijling, and J. Ma, "Forecasting daily streamflow using hybrid ann models," *Journal of Hydrology*, vol. 324, no. 1, pp. 383–399, 2006.

[39] A. Roy, L. S. Kim, and S. Mukhopadhyay, "A polynomial time algorithm for the construction and training of a class of multilayer perceptrons," *Neural Networks*, vol. 6, no. 4, pp. 535–545, 1993.

[40] J. Sietsma and R. J. F. Dow, "Neural net pruning-why and how," in *Proceedings of the IEEE International Conference on Neural Networks*, pp. 325–333, IEEE, 1988.

[41] Z. Wang, C. Di Massimo, M. T. Tham, and A. Julian Morris, "A procedure for determining the topology of multilayer feedforward neural networks," *Neural Networks*, vol. 7, no. 2, pp. 291–300, 1994.

[42] K. G. Sheela and S. N. Deepa, "Selection of number of hidden neurons in neural networks in renewable energy systems," *Journal of Scientific & Industrial Research*, vol. 73, no. 10, pp. 686–688, 2014.

[43] M. M. Fischer and S. Gopal, "Artificial neural networks: a new approach to modeling interregional telecommunication flows," *Journal of Regional Science*, vol. 34, no. 4, pp. 503–527, 1994.

[44] J. L. Lipae and E. P. Deligero, "On forecasting water consumption in davao city using autoregressive integrated moving average (arima) models and the multilayer perceptron neural network (mlpnn) processes," *International Journal of Humanities and Applied Sciences*, vol. 1, no. 4, pp. 2277–4386, 2012.

[45] R. Š. Babić, I. Grgurević, B. Eng, and Z. Majić, "Comparison of air travel demand forecasting methods," in *Proceedings of the 14th International Conference on Transport Science Maritime, Transport and Logistics Science (ICTS '11)*, 2011.

[46] A. Kadiyala and A. Kumar, "Univariate time series prediction of air quality inside a public transportation bus using available software," *Environmental Progress & Sustainable Energy*, vol. 31, no. 4, pp. 494–499, 2012.

[47] L. Prechelt, "Automatic early stopping using cross validation: Quantifying the criteria," *Neural Networks*, vol. 11, no. 4, pp. 761–767, 1998.

[48] C. C. Gowda and S. Mayya, "Comparison of back propagation neural network and genetic algorithm neural network for stream flow prediction," *Journal of Computational Environmental Sciences*, 2014.

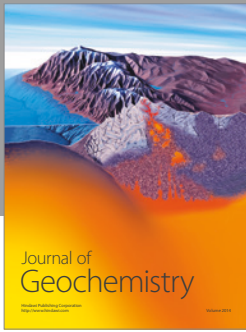
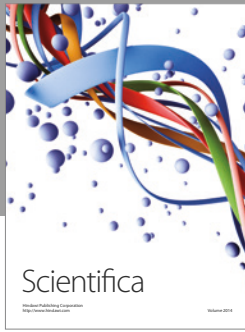
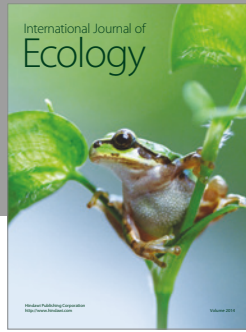
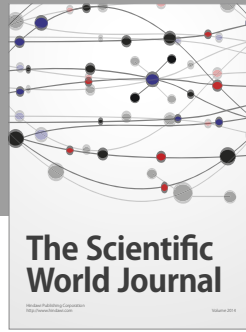
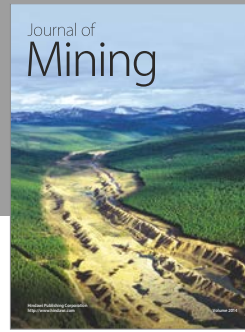
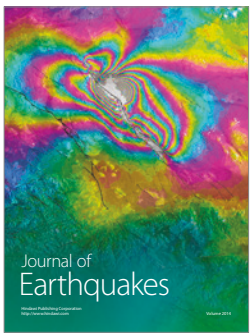
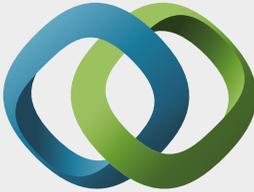
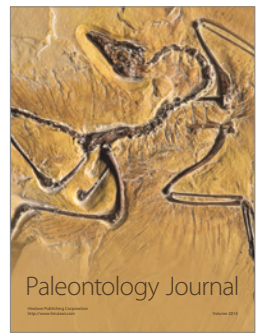
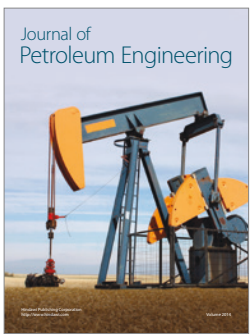
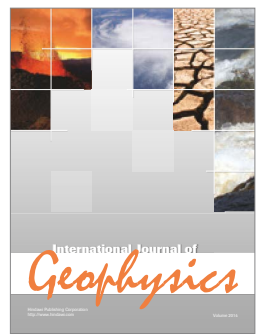
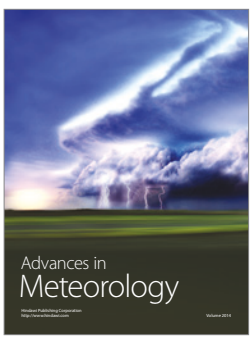
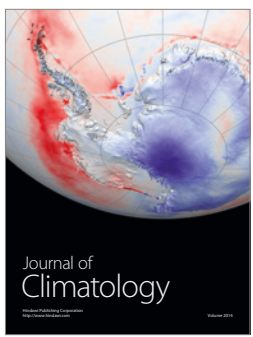
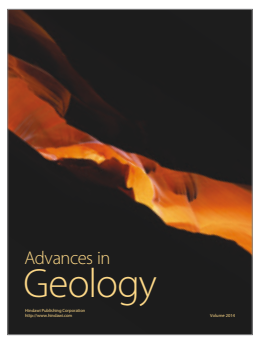
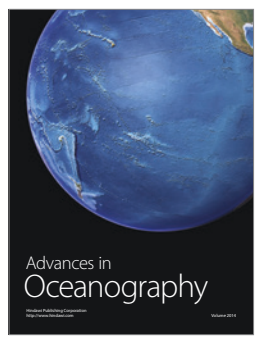
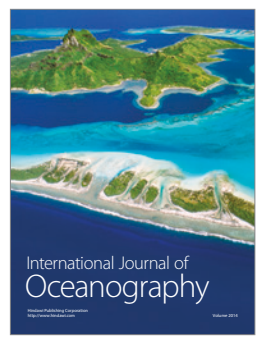
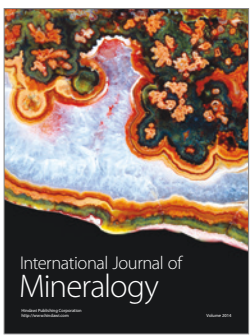
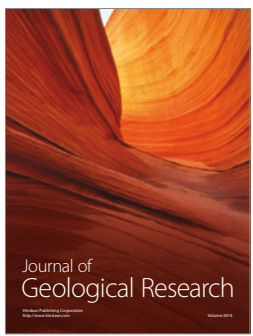
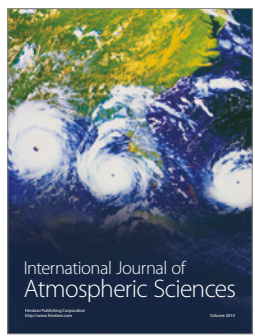
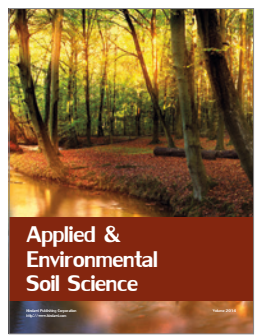
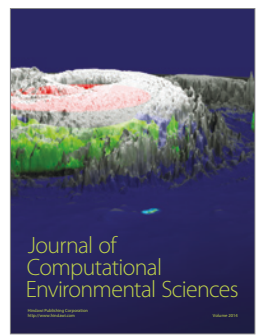

Submit your manuscripts at
https://www.hindawi.com